# Effects of mass loss for highly-irradiated giant planets

W. B. Hubbard[*], M. F. Hattori[*], A. Burrows[†], I. Hubeny[†], and D. Sudarsky[†]

[*]Lunar and Planetary Laboratory, University of Arizona, Tucson AZ 85721, USA
[†]Department of Astronomy, University of Arizona, Tucson AZ 85721, USA

**Abstract**

We present calculations for the evolution and surviving mass of highly-irradiated extrasolar giant planets (EGPs) at orbital semimajor axes ranging from 0.023 to 0.057 AU using a generalized scaled theory for mass loss, together with new surface-condition grids for hot EGPs and a consistent treatment of tidal truncation. Theoretical estimates for the rate of energy-limited hydrogen escape from giant-planet atmospheres differ by two orders of magnitude, when one holds planetary mass, composition, and irradiation constant. Baraffe *et al.* (2004, *A&A* 419, L13-L16) predict the highest rate, based on the theory of Lammer *et al.* (2003, *Astrophys. J.* 598, L121-L124). Scaling the theory of Watson *et al.* (1981, *Icarus* 48, 150-166) to parameters for a highly-irradiated exoplanet, we find an escape rate ~$10^2$ lower than Baraffe's. With the scaled Watson theory we find modest mass loss, occurring early in the history of a hot EGP. In this theory, mass loss including the effect of Roche-lobe overflow becomes significant primarily for masses below a Saturn mass, for semimajor axes ≥ 0.023 AU. This contrasts with the Baraffe model, where hot EGPs are claimed to be remnants of much more massive bodies, originally several times Jupiter and still losing substantial mass fractions at present.

## 1. Introduction

At present considerable uncertainty exists as to whether extrasolar giant planets (EGPs) are likely to suffer appreciable mass loss over their lifetime when subjected to high levels of XUV irradiation from their host star. A major question concerning the origin of short-period (period $P \sim$ days, semimajor axis $a \sim$ few $\times 10^{-2}$ AU) EGPs concerns this point: do short-period (hot) EGPs originate with their presently-observed masses at $a \sim 10$ AU and migrate rapidly to $a \sim$ few $\times 10^{-2}$ AU without appreciable mass loss, or were the original bodies several times more massive, suffering continual rapid mass loss throughout their evolution? The latter scenario, advocated by Baraffe *et al.* (2004), implies that the observed hot EGPs are remnants of much more massive objects and are continuing to lose mass at a significant rate at present. However, as we show in this paper, a theory based on the results of Watson *et al.* (1981) implies that only hydrogen-rich EGPs with mass $M < 0.2\, M_J$ (where $M_J$ = Jupiter's mass) will be significantly affected by mass loss.

We do not present here any independent calculations of the escape rate $\Phi$ (molecules cm$^{-2}$ s$^{-1}$) as a function of $Q_0$ (erg cm$^{-3}$ s$^{-1}$), the volume-heating rate of the planetary atmosphere due to absorption of stellar XUV radiation. For energy-limited escape, the two quantities are proportional. As we discuss below, for fixed XUV irradiation from a solar-type star, the Watson theory gives a proportionality constant, and therefore an atmospheric escape rate, that is $10^{-2}$ of the escape rate predicted by Baraffe *et al.* Tidal effects have been extended to the Lammer formulation (Jaritz *et al.* 2005), but not explicitly in the Watson theory. A calculation by Yelle (2004) originally gave an escape rate about two orders of magnitude lower than the Watson rate, but after taking into account the corrigendum by Yelle (2006), we find that the revised Yelle escape rate is no longer an outlier but rather lies between the Baraffe rate and the scaled Watson rate. Thus, the purpose of the present paper is to present, side by side, the implications of the two extreme theories (Baraffe and Watson), by consistently including tidal effects in energy-limited escape for both.

## 2. Scaling Relations

### *2.1. Effect of tides on atmospheric binding*

The critical parameter for the binding of the EGP atmosphere is usually written as (Yelle 2004)

$$\lambda = \frac{GMm}{kTr}, \tag{1}$$

where $G$ is the gravitational constant, $m$ is the molecular mass (2 amu for $H_2$), $k$ is Boltzmann's constant, $T$ is the atmospheric temperature, and $r$ is the atmospheric radius. For a planet at great distance from its primary, one may write

$$\lambda = \frac{mV_0}{kT}, \tag{2}$$

where $V_0$ is the gravitational potential of the atmospheric layer and $E_B = mV_0$ is the gravitational binding energy of a molecule. In energy-limited escape, a certain fraction of the energy from the star impinging on the EGP atmosphere, averaged over the stellar spectrum, removes a hydrogen molecule from the atmosphere by supplying it with kinetic energy > $mV$. The efficiency of this process is much less than unity because only XUV photons are effective in heating the upper atmosphere. The efficiency is reduced further because part of the XUV photon energy drives dissociation, ionization, and other reactions without removing the reaction products from the atmosphere. Let the efficiency of this process be

$$\varepsilon = \frac{\Phi E_B}{S_*}, \tag{3}$$

where $S_*$ is the "solar" (or stellar) constant, the photon power normally incident on the EGP atmosphere at the planet's orbital radius $a$. The predicted efficiency factor $\varepsilon$ ranges from $\sim 10^{-4}$ for the Baraffe model (meaning that for every $10^4$ ergs of stellar photon energy averaged over wavelength that impinge on the EGP atmosphere, ~ 1 erg is available to remove a hydrogen molecule) to $\sim 10^{-6}$ for the scaled Watson model.

The above arguments require modification when a significant tidal potential is present, i.e., when $r$ is only a factor ~2 to 3 smaller than $r_H$, where $r_H$ is the radius of the planet's Hill sphere. For example, this will be the case for an EGP orbit about a 1 $M_\odot$ (solar mass) star with a semimajor axis $a \sim 0.023$ AU. When this is the case, we recognize that the gravitational binding energy of a molecule is no longer $E_B = mV(r)$, but rather

$$E_{B,\text{mod}} = mV_{\text{mod}}(r) - mV_{\text{mod}}(L1) = m\Delta V_{\text{mod}}, \tag{4}$$

where $V_{\text{mod}}$ is the usual modified gravitational potential in the planet's frame, including the stellar and planetary gravitational potentials and the effective rotational potential. At the L1 point between the planet and star (as well as at the nearly-symmetric L2 on the other side of the planet), molecules are not gravitationally bound to the planet and can go into independent orbit about the star. Strictly speaking, when $r$ is very close to $r_H$, effectively close to L1 or L2, the planet is strongly distorted from a spherical shape and the modified potential should include this distortion. However, as this limit is approached, mass loss occurs very quickly, by virtue of a modified binding parameter,

$$\lambda_{\text{mod}} = \frac{E_{B,\text{mod}}}{kT} \to 0, \tag{5}$$

and imprecision in the evaluation of $E_{B,\text{mod}}$ will affect the mass limit only slightly.

In summary, the effect of tides is represented in the scaled theory by multiplying the escape rate $\Phi(Q_0)$ by the tidal enhancement factor $\tau = V_0/\Delta V_{\text{mod}} > 1$. Figure 1 gives an example of the calculation of the tidal enhancement factor for EGPs at an orbital radius $a = 0.023$ AU, equal to that of OGLE-TR-56b (whose mass $M = 1.45 \pm 0.23$ $M_J$; Bouchy *et al.* 2005), for a 1-$M_\odot$ primary star (Note that the actual mass of OGLE-TR-56 is $1.04 \pm 0.05\,M_\odot$). The orbital radius of 0.023 AU corresponds to the inner radius limit of detected EGPs and is, evidently, the distance at which mass loss processes will be most pronounced. Figure 1 illustrates the point that the tidal limit is of course a function of the planetary radius $r$, which in turn is a function of the internal entropy of the EGP. The solid curve represents the relative reduction in the binding energy for an internal entropy corresponding to an isolated cooling age of 1 Ma, and the dotted curve corresponds to an age of 5 Ma. The curves osculate the abscissa at $M \sim 0.15$ $M_J$, which corresponds roughly to the point at which the planet would fill its Roche lobe; the atmosphere is completely unbound at this point. The curves turn back up for smaller masses, but there is no physical significance to this branch. Further discussion of the tidal limit is presented in Section 3.

*2.2. Description of models*

Watson *et al.* (1981; hereafter model W) consider mass loss from a planet's hydrogen exosphere at 1 AU from a solar-type star (specifically, the Earth). Model W introduces the quantity $S$ (erg cm$^{-2}$ s$^{-1}$), the sphere-averaged, efficiency-corrected areal heating rate in the planetary atmosphere due to XUV solar irradiation, such that

$$S = \int dh Q_o(h) . \tag{6}$$

Here the integral is carried out over the height range $h$ in the planetary thermosphere where $Q_0(h)$ is appreciable. Watson *et al.* find $S \sim 1$ erg cm$^{-2}$ s$^{-1}$, to be compared with the solar constant at Earth, $1.4 \times 10^6$ erg cm$^{-2}$ s$^{-1}$. Thus, the efficiency of XUV heating having already been incorporated in the definition of $S$, model W predicts $\varepsilon \sim 10^{-6}$, for a planet at 1 AU from the Sun, where tidal effects are negligible. This efficiency factor can also be deduced by substituting the model W value for the Earth, $\Phi_W \sim 2 \times 10^{11}$ molecules cm$^{-2}$ s$^{-1}$, in Eq. (3). We will employ this result, valid for a hydrogen exosphere at 1 AU from a 1-$M_\odot$ dwarf at an age of 4.5 Ga, to scale the results of model W to highly-irradiated EGPs. A recent model by Tian *et al.* (2005), obtains EGP mass-loss rates that are within a factor of a few of those predicted by model W, as does the revised Yelle (2006) model.

Baraffe *et al.* (2004; hereafter model B) calculate a quantity (which they call $F$*) equivalent to model W's $S$ and obtain essentially the same value for Earth-like parameters: $S \sim 2$ erg cm$^{-2}$ s$^{-1}$. However, their calculated loss rate $\Phi_B$ is $\sim 10^2\ \Phi_W$, where $\Phi_W$ is the model W loss rate for the same irradiation parameters. Thus we scale model B to model W by setting the efficiency factor for model B: $\varepsilon_B \sim 10^2\ \varepsilon_W \sim 10^{-4}$. Yelle (2004, 2006) notes that $\Phi$ is almost precisely proportional to $a^{-2}$, as would be expected for energy-limited escape.

In the following discussion, we calculate mass-loss histories for the two extreme models, B and W, writing

$$\Phi_B = \tau(t) \frac{\Phi_W}{a^2} \frac{S(t)}{S(4.5\ \mathrm{Ga})} \frac{\varepsilon_B}{\varepsilon_W} . \tag{7}$$

In Eq. (7), $a$ is in AU, and $S(t)$ is scaled to the value for the present Sun at age $t = 4.5$ Ga. In Eq. (7), $\tau(t)$ is calculated using the time-dependent mass and radius of the EGP. Although the masses of the host stars of known highly-irradiated EGPs differ by modest factors from 1 $M_\odot$, for the purposes of this paper we approximate all of these stars by the Sun. Particularly for the B model, the early history of $S(t)$ is important, and we approximate it by

$$\frac{S(t)}{S(4.5\ \mathrm{Ga})} = \left( \frac{4.5\ \mathrm{Ga}}{t} \right)^{1.23} \tag{8}$$

(Ribas *et al.*, 2005). This expression has a singularity at $t = 0$, meaning that there is a spike in mass loss at early epochs. However, the mass loss rate has no meaning at times earlier than the presumed formation time of the EGP at $t = 1$ or 5 Ma, so Eq. (1) is cut off at this starting point. Under this assumption, the EGP mass range where significant mass loss occurs is not strongly affected by the model for the early history of $S(t)$, for the W model. In contrast, for a broad range of masses in the B model, a large fraction of the cumulative mass loss occurs when $t < 1$ Ga.

## 3. Coupling of mass loss models to interior evolution

We have prepared a grid of surface conditions for EGP evolution at four roughly equally spaced, representative distances from the primary star, namely, 0.023 AU (OGLE-TR-56b), 0.034 AU, 0.046 AU (HD209458b; Brown *et al.* 2001, Schneider 2005), and 0.057 AU. The orbital radius $a$ = 0.034 AU was chosen to correspond to a value midway between the orbits of OGLE-TR-56b and HD209458b. The orbital radius $a$ = 0.057 AU was chosen to be spaced at the same interval as the previous three. It does not coincide with any particular EGP.

The method for computation of these grids is described in Burrows *et al.* (2004). In all cases the primary star is assumed to be the present-day Sun, a G2V dwarf at an age of 4.5 Ga. We recognize that this assumption introduces an inconsistency with Eq. (7), which means that the limits for stability of highly-irradiated EGPs derived below will be affected to the extent that the mass loss at ages < 1 Ga is important. In most cases where this is true, other uncertainties, to be described below, are dominant. Where ages of the EGP host stars are known, those ages range from ~ 3 to 10 Ga, with a mode value of 3.3 Ga (Takeda *et al.* 2006). At a fixed distance to the star, the surface-condition grid is a table giving the asymptotic (deep interior) specific entropy, $S/Nk$ as a function of the surface gravity $g$ and the effective temperature $T_{\rm eff}$. The EGP's "surface" is defined as the 1-bar level, and $T_{\rm eff}$ characterizes the *deep* (intrinsic) heat flux, such that the latter is given by $T_{\rm eff}^4$ multiplied by the Stefan-Boltzmann constant, $\sigma$. Here $S/Nk$ is the entropy per baryon per Boltzmann's constant, computed using a solar-composition mixture of hydrogen and helium with the Saumon *et al.* equation of state (1995). When referring to the specific entropy, we always use the quantity $S/Nk$, which should not be confused with the areal heating rates $S$ or $S_*$. Moreover, the deposition of XUV energy, characterized by $S$, occurs high in the EGP atmosphere, affecting only the mass-loss rate. The bulk of the irradiation energy (which scales with $S_*$, >> $S$) is deposited much deeper in the atmosphere and is incorporated, via the surface-condition grid, in $S/Nk(g,T_{\rm eff})$.

One of the main objectives of this paper is to define the lower limit for stability to mass loss as a function of $a$. For the W model, this limit is primarily determined by the tidal stripping criterion, i.e., the planetary radius $r$ as compared with $r_H$. In the low mass range, close to the stability limit, one has $r \sim r_H$, but $r$ is an increasing function of $S/Nk$. As the planet radiates deep-interior heat to space, $S/Nk$ decreases. Thus the stability limit, basically determined by $\tau \rightarrow \infty$, is a function of the planet's initial $S/Nk$. It therefore turns out that the stability limit (for the W model) depends on the initial formation history of the EGP. To address this problem in a very approximate way, we assume that the EGP is initially formed during the first $10^6$ years of the system's existence, at an initial orbital radius > 1 to 10 AU, where irradiation and tidal effects are negligible. We then assume that such an EGP subsequently migrates inward under the influence of torques from the protostellar nebula (Lin *et al.* 2000) and reaches its final orbital radius $a$ by a system age $t \sim 1$ to 5 Ma, at which point the nebula and resulting torques have vanished. Note, however, that this scenario has been questioned by Wuchterl (2001).

The initial $S/Nk$ of the EGP is estimated by computing the approximate $S/Nk$ of an isolated (non-irradiated) EGP of mass $M$ at the specified age $t$. The implications of this calculation for the tidal mass limit were shown in Fig. 1, where the solid curve shows $\tau(M)$ for $S/Nk$ corresponding to an age $t$ = 1 Ma and the dotted curve corresponds to $t$ = 5 Ma. This calculation is not intended to define the tidal stability limit with great exactitude, nor does it address the question of whether a solar-composition mass < 0.2 $M_J$ can be formed in the first place.

Obviously, the tidal enhancement factor becomes even less pronounced for the models at larger values of $a$. Figure 2 shows $\tau$ at age $t$ = 1 Ma and three orbital radii (including $a$ = 0.046 AU, the orbital radius of HD 209458b, but for a 1-$M_\odot$ primary; the actual mass of HD 209458 is 1.05 $M_\odot$).

## 4. Results

### *4.1. Boundary conditions*

We generated surface-condition grids $S/Nk(g,T_{\rm eff})$ for the four orbital radii, as calculated for a 1-$M_\odot$ primary at $t$ = 4.5 Ga. The model-atmosphere grids were constructed by using ~ 10 equally-spaced values of log $g$ (cm $s^{-2}$) and $T_{\rm eff}$ (K) as independent variables and solving for the deep entropy. The resulting ~ 100 data points were then fitted with a smooth surface. A small amount of extrapolation was necessary for the most extreme B models. In the case of low-$g$ models (log $g$ < 1.5) , the assumption of a plane-parallel atmosphere becomes suspect; this problem primarily arises for low-mass models ($M$ << 1 $M_J$) close to the stability limit. Since no calculations were performed for $T_{\rm eff}$ < 50 K, the $S/Nk$ surface was truncated at very low values.

### *4.2. Stability limit*

The term, "stability limit," requires some discussion, since the physical mechanism for the non-persistence of EGPs due to mass loss varies with orbital radius $a$ and efficiency factor $\varepsilon$. In general, our calculations do not exhibit a dynamical instability whereby a perturbation to the EGP, such as removal of an atmospheric mass layer, is self-accelerated, growing exponentially with time. A classical example of such behavior would be a gaseous sphere with a polytropic index greater than one (Chandrasekhar 1939), filling its Roche lobe in proximity to the primary star. For such an object, a decrease in mass $M$ results in an increase of radius $R$ and overflow of the Roche lobe at an accelerating rate. However, EGPs with $M$ ~ 1 $M_J$ have effective polytropic indices close to unity, with almost no variation of radius with mass. Thus, this instability can only appear for hydrogen-rich bodies of Jovian-like entropy, but with $M$ << 1 $M_J$, when the pressure-density relation is such that $R$ increases with decreasing $M$. For the W models, Roche-lobe overflow is the operative stability limit, and we do observe an abrupt instability, usually within mass and time ranges too fine to resolve, for masses where $\tau$ becomes very large, as discussed in Section 2.1. For the B models, $\varepsilon$ is so large that masses $M$ ~ 1 $M_J$ suffer continual mass loss without exhibiting a dynamical instability. For the latter models, there is thus, strictly speaking, no mass limit defined by stability considerations. Our compromise definition of "stability limit," which can be applied to both mass-loss models, is that the limiting mass at orbital radius $a$ is the initial mass which will decrease by at least a factor two over a time period of 5 Ga.

Evolution begins at the $S/Nk$ value where the EGP of mass $M$ is deposited at an age of 1 or 5 Ma. Models initially deflect to lower surface gravity, caused by the high initial XUV flux from the young host star and corresponding high mass loss, as given by Eq. (8). The EGP evolves downward in entropy to the final evolution point, which corresponds to $t = 5$ Ga or $M = 0$, whichever occurs first. For $a = 0.023$ AU, the initial masses that decrease by a factor of two after 5 Ga are 4.6 $M_J$ (model B) and 0.82 $M_J$ (model W). The result for the B model is generally consistent with the results presented by Baraffe *et al.* (2004). However, the critical mass range for model B is so large that we must modestly extrapolate our $S/Nk(g,T_{\rm eff})$ surface to larger values of $g$.

For the two most massive W models (0.35 and 0.40 $M_J$), mass loss is so slow that the models show increasing surface gravity as a result of contraction, despite mass loss, while the 0.30 $M_J$ W model sheds mass at a sufficiently rapid rate to disappear within 5 Ga. Thus, in the case of the W model at $a = 0.023$ AU, one Saturn mass is at the stability limit. The B models all show decreasing surface gravity with time as mass loss is substantial throughout their evolution.

### *4.3. Mass loss and radius variations*

In Fig. 3, we plot the $M(t)$ trajectories for B models at the stability limit. These curves correspond to an initial mass of 4.6 $M_J$ at 0.023 AU, 3.2 $M_J$ at 0.034 AU, 2.7 $M_J$ at 0.046 AU, and 2.3 $M_J$ at 0.057 AU. The cusp in $M(t)$ as $t \rightarrow 0$ is a consequence of Eq. (8). Figure 4 shows the evolution of radius with time, for mass-loss model B and for the same initial masses shown in Fig. 3.

Figure 5 shows $M(t)$ trajectories for the W models. These curves correspond to initial masses of 0.82 $M_J$ at 0.023 AU, 0.56 $M_J$ at 0.034 AU, 0.44 $M_J$ at 0.046 AU, and 0.37 $M_J$ at 0.057 AU. Note that the ordinate range is much smaller than in Fig. 3. As in the case of the B model, there is a cusp in the early M(t) curves due to high initial XUV irradiation.

Figure 6 shows the evolution of radius with time, for the same models plotted in Fig. 5 (mass-loss model W).

As has already been emphasized by Baraffe *et al.* (2004), if mass-loss model B is correct, HD 209458b would have to be a remnant of a substantially more massive object. For the W model at the distance of HD 209458b, rapid cooling causes a rapid increase of surface gravity despite mass loss, and the marginal model is only ~50% more massive than Neptune (Neptune's mass $M = 0.05$ $M_J$). Determination of the radius of a Neptune-mass EGP might help to resolve whether predominantly-hydrogen objects of such low mass exist. However, the absence of an example of a Neptune-mass hydrogen planet might not be so much a test of mass-loss theories as a constraint on formation mechanisms for small hydrogen objects.

As we see, the B model implies that only planets with an initial mass greater than 2.3 $M_J$ can persist for 5 Ga as an EGP at $a = 0.057$ AU, but only marginally, and with substantially-reduced mass at 5 Ga. For the W model, the stability limit for the initial mass at this distance is ~ 1 Saturn mass.

## 5. Discussion

Figure 7 and Table 1 summarize the principal results of our calculations. In Fig. 7, the ordinate represents the initial mass of an EGP whose mass will be reduced by a factor of ~2 after 5 Ga. For the B models, the limit is primarily established by the very large value $\varepsilon_B \sim 10^{-4}$. The limit for the W models is also affected by tidal enhancement of the mass loss rate. Also shown in Fig. 7 are labeled data points for transiting EGPs, as well as lower mass limits for highly-irradiated EGPs of unknown orbital inclination (Schneider 2006, Rivera *et al.* 2005, Bouchy *et al.* 2005, Bouchy *et al.* 2004, Butler *et al.* 2004, Bouchy *et al.* 2004, Moutou *et al.* 2005, Udry *et al.* 2003, McArthur *et al.* 2004, Alonso *et al.* 2004, Bouchy *et al.* 2005, Marcy *et al.* 2000, Sato *et al.* 2005, Brown *et al.* 2001, Pont *et al.* 2005, Butler *et al.* 2003).

The initial-mass limits shown in Fig. 7 are conveniently represented by the following polynomial fit:

$$\log M = c_0 + c_1 a + c_2 a^2, \quad (9)$$

where the initial mass $M$ is in $M_J$ and the orbital radius $a$ is in AU. Values for the coefficients for the two mass-loss models are given in Table 1. Eq. (9) is valid for the range of $a$ shown in Fig. 7, i.e. ~ 0.02 to 0.06 AU, for a solar-mass primary star.

Comparison of these data points with our results from applying the two theories indicates that if the extreme B model were correct, virtually all observed hot EGPs must be remnants of much more massive bodies, originally several times Jupiter and still losing substantial mass fractions at present. Because the mass-loss rate in the B theory is so extreme, a large fraction of the cumulative mass loss occurs for $t \ll 1$ Ga, when the young primary star has a large XUV flux.

Implications of the W theory are very different (see Fig. 7): observed hot EGPs retain practically all of the nebular-gas mass fraction with which they were formed. The W theory predicts that Saturn-mass, hydrogen-dominated EGPs would not persist at orbital radii less than about 0.05 AU (for solar-type primaries), while such objects could exist were the mass-loss rate even lower than this limiting model. If the B theory is correct, Saturn-mass EGPs within ~ 0.1 AU of their primaries could exist only for a brief interval on their way to complete evaporation. The two mass-loss theories have radically different implications for the mass distribution statistics of highly-irradiated EGPs (Hubbard and Hattori, in preparation).

The observations by Vidal-Madjar (2003) of an extended hydrogen atmosphere around HD209458b provide an observational constraint on Φ for this object, but not a direct determination. As discussed by Yelle (2004, 2006), Vidal-Madjar's interpretations of their observations imply a mass-loss flux in the same range as model W.

For even the most conservative mass loss theory, the ~ Neptune-mass planets Gliese 876d, GJ 436b, and 55 Cnc e are unlikely to be composed primarily of hydrogen. This conclusion must be confirmed by further calculations, however, because Gliese 876 and GJ 436 are M dwarfs rather than G dwarfs, so their EGP mass flux due to irradiation must be scaled accordingly.

**Acknowledgments**
This research was supported by NASA Grant NAG5-13775 (PGG) and NASA Grants NNG04GL22G and NNG05GG05G (ATP).

**Tables**

Table 1
Values of coefficients to be used in Eq. (9)

| Model | $c_0$ | $c_1$ | $c_2$ |
|---|---|---|---|
| Baraffe | 1.10 | −24.9 | 198 |
| Watson | 0.43 | −27.9 | 218 |

**Figures and Captions**

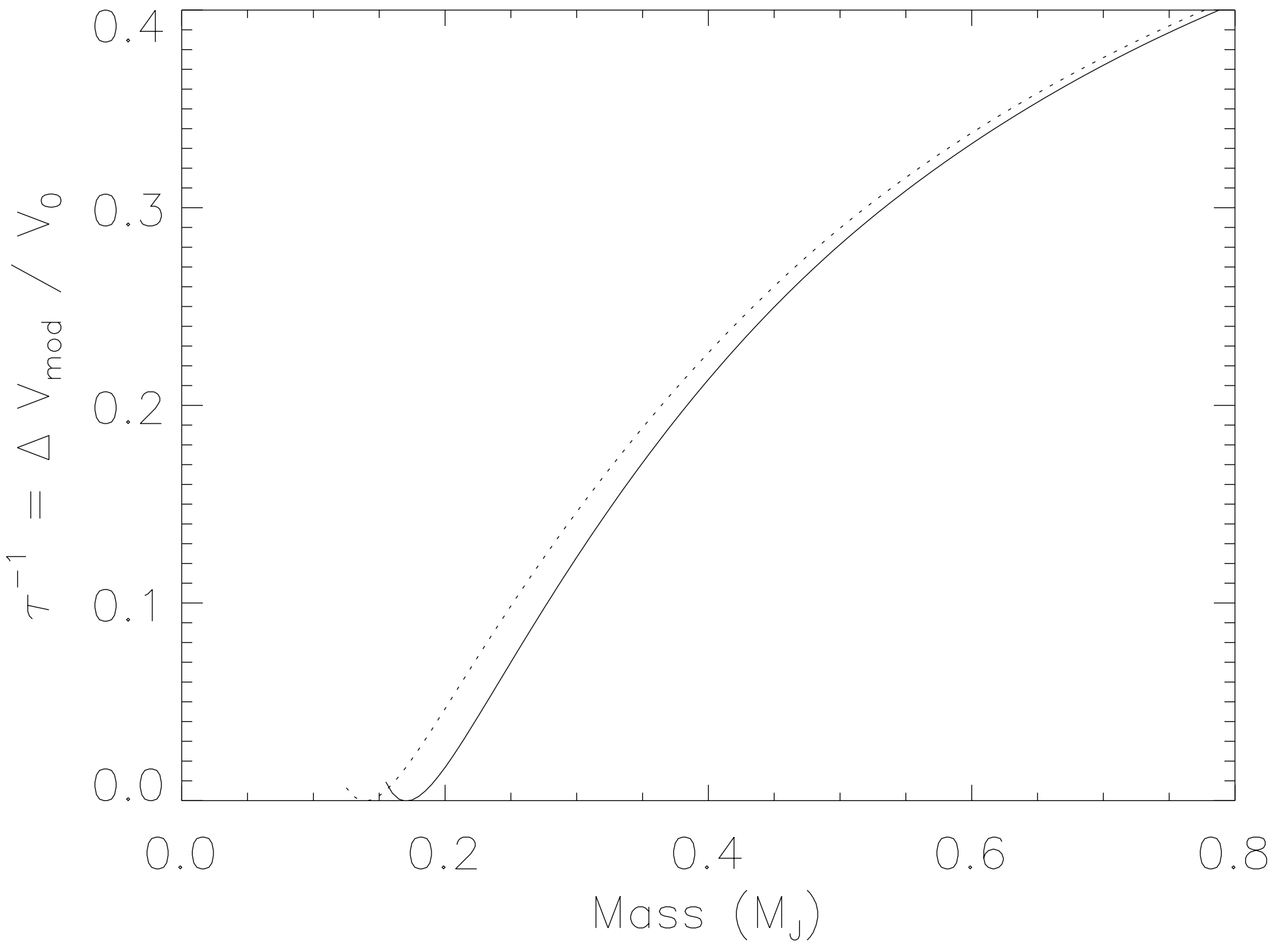

**Fig. 1** – Reciprocal of the tidal enhancement factor for masses close to the tidal limiting mass (the mass for which the ordinate goes to zero), for a 1-$M_{\odot}$ star and *a* = 0.023 AU (similar to OGLE-TR-56b). See text for further explanation.

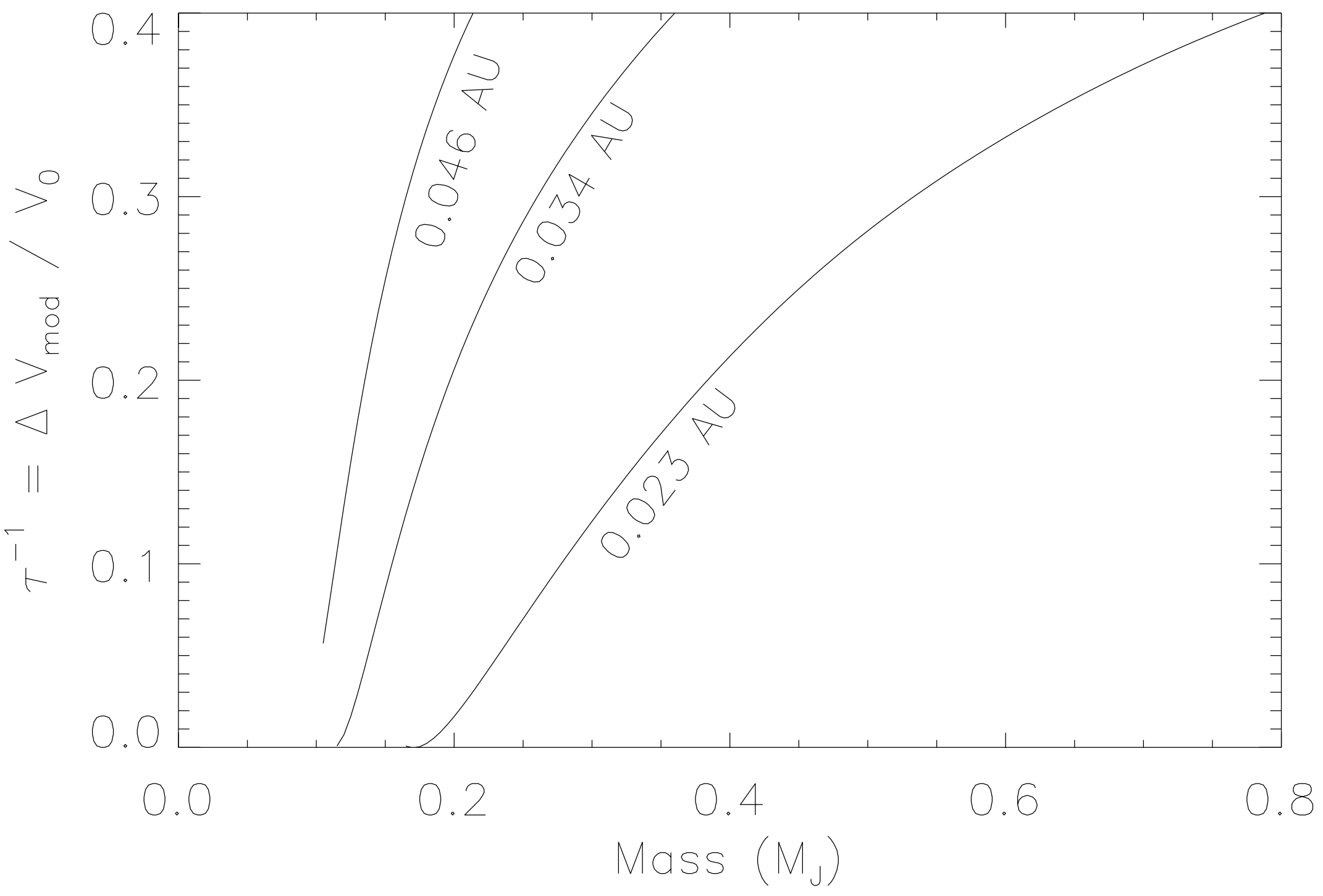

**Fig. 2** – Reciprocal of the tidal enhancement factor at 1 Ma for a 1-$M_\odot$ star and three orbital radii.

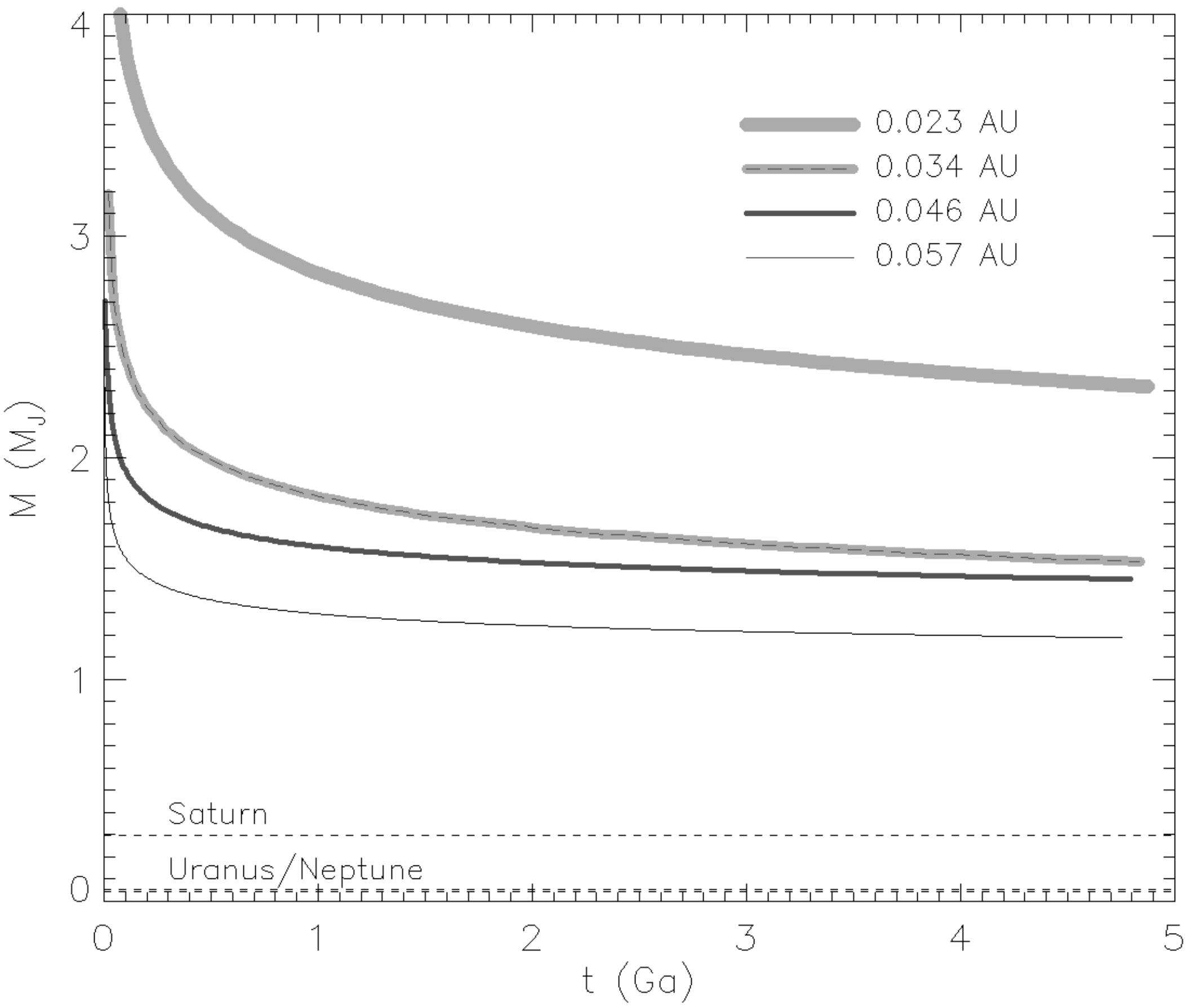

**Fig. 3 –** Model B predictions of mass with time for EGPs suffering 50 percent mass loss over 5 Ga, for the four values of *a*. The thickest lines correspond to the hottest EGPs at *a* = 0.023 AU, while the thinnest lines correspond to *a* = 0.057 AU. Horizontal dashed lines are at the masses of Saturn and of Neptune and Uranus.

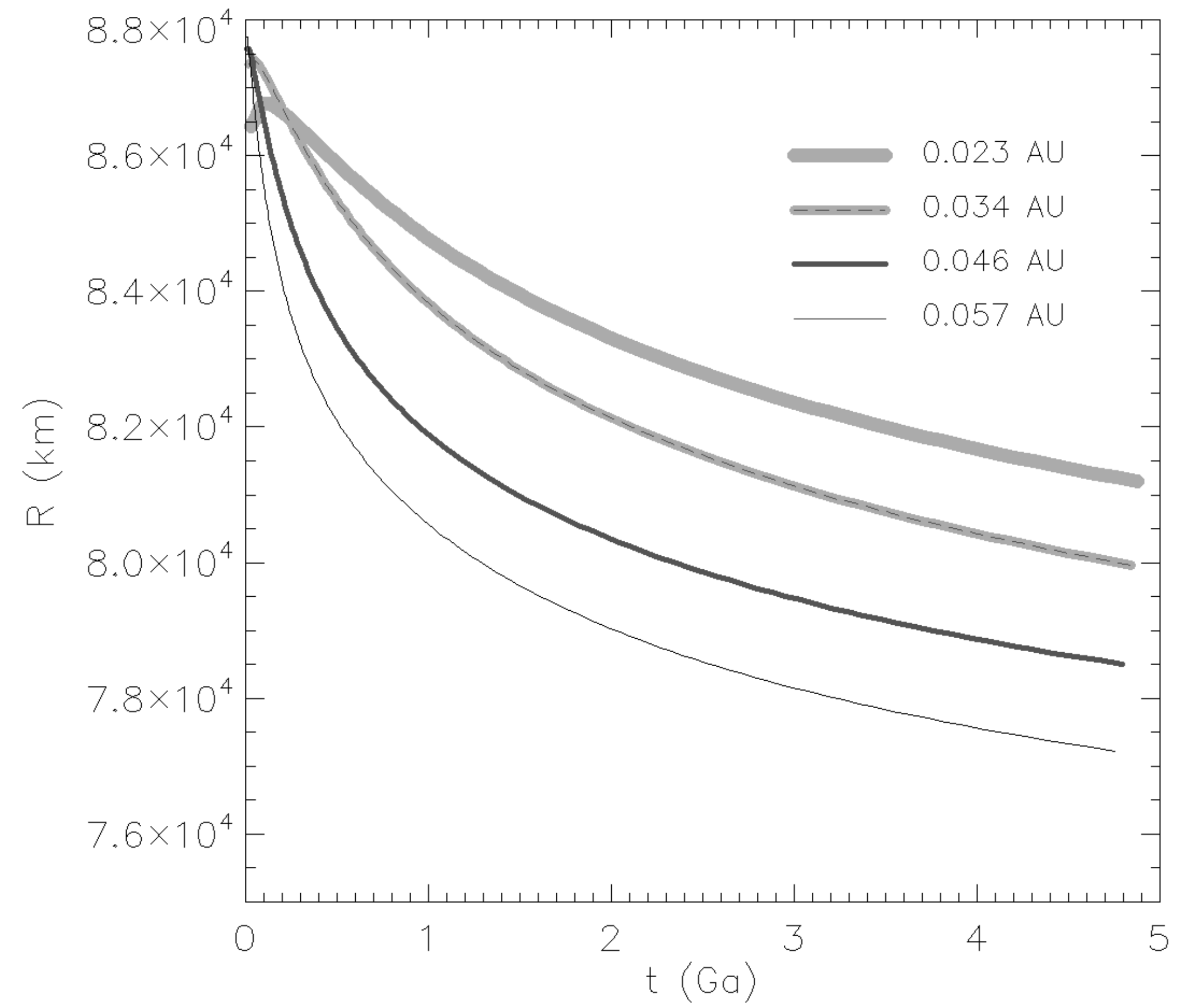

**Fig. 4** – Model B predictions of radius with time for EGPs suffering 50 percent mass loss over 5 Ga, for the four values of $a$.

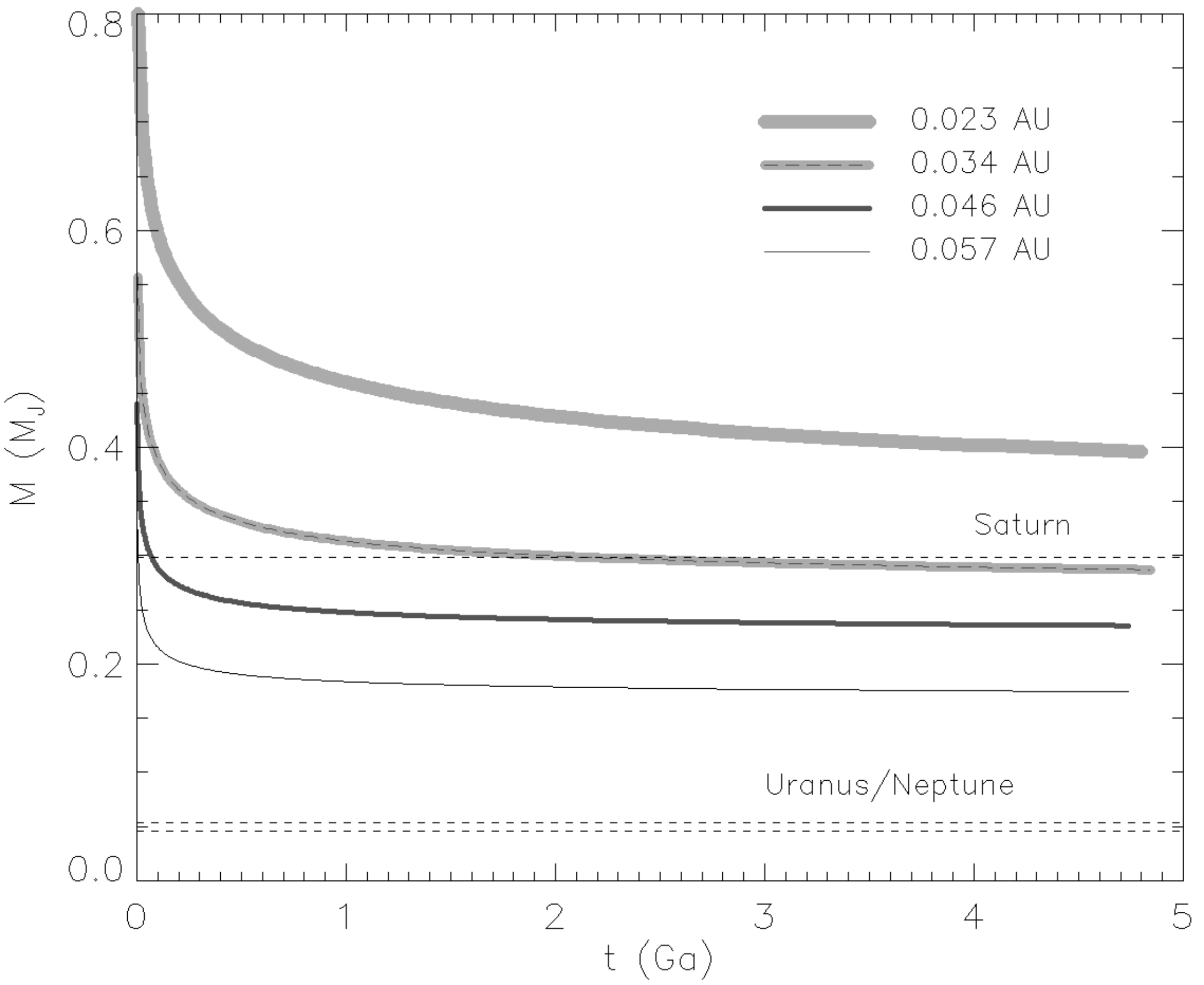

**Fig. 5 –** Model W predictions of mass with time for EGPs suffering 50 percent mass loss over 5 Ga, for the four values of *a*.

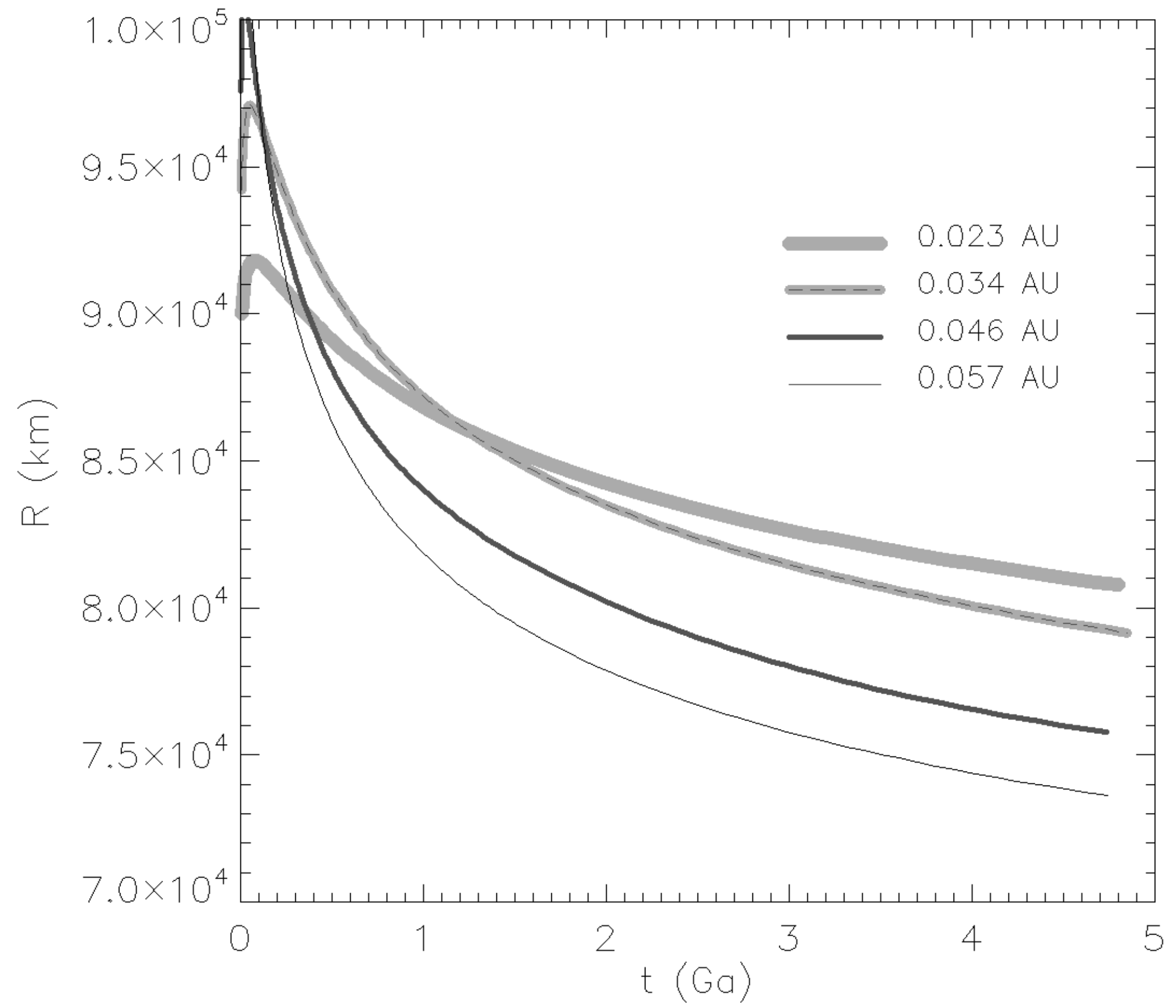

**Fig. 6 –** Model W predictions of radius with time for EGPs suffering 50 percent mass loss over 5 Ga, for the four values of *a*. The curves show the effects of initial high mass loss, per Eq. (8).

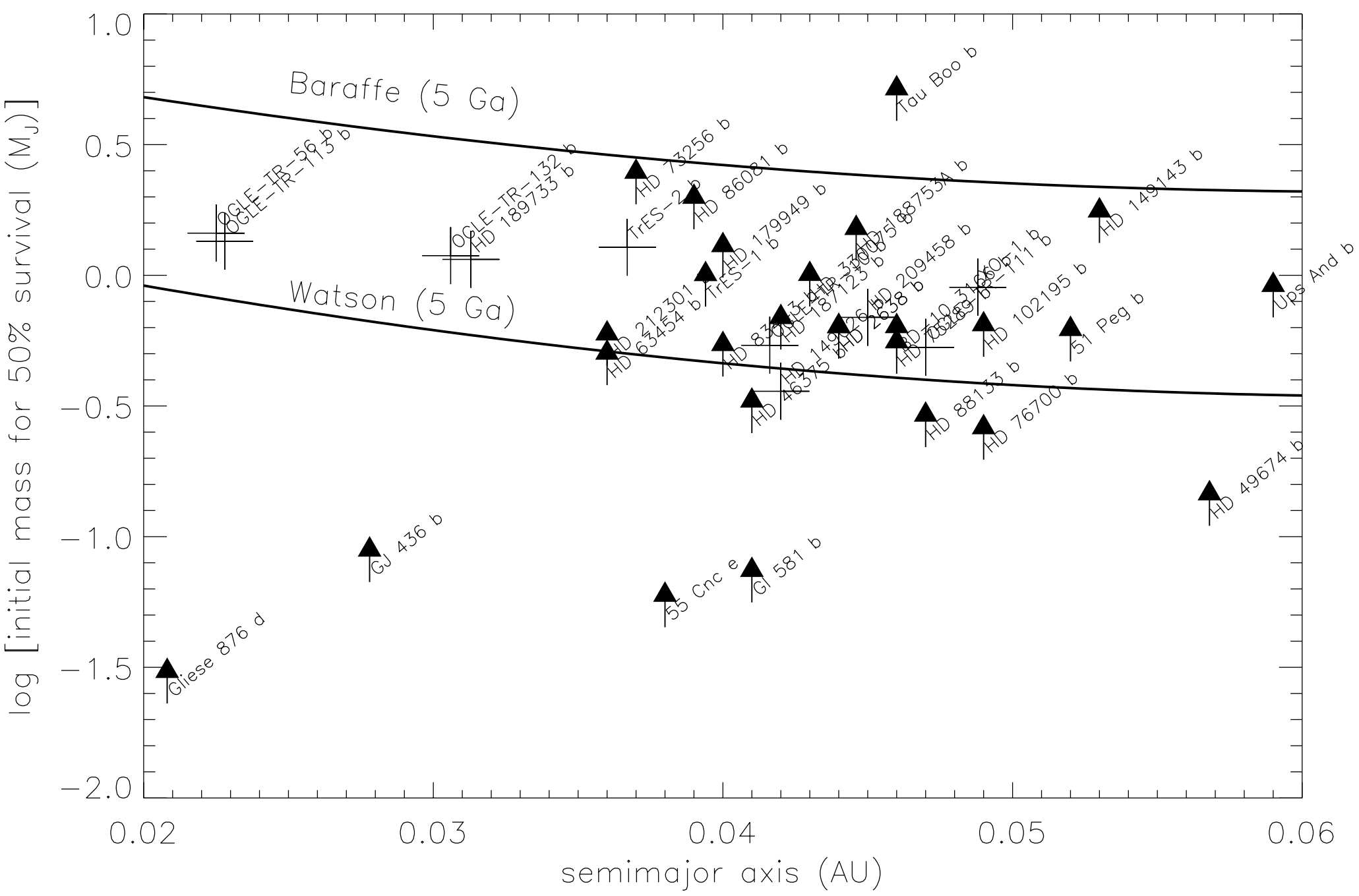

**Fig. 7** – Heavy black curves show the limiting mass for stability of EGPs orbiting a solar-type star at various orbital distances, for the three mass-loss models. Observed EGPs are plotted for comparison, as pluses (where mass is known), or as arrows showing the lower limit on mass. In the cases of Gl 876 d and GJ 436 b, the primary has a mass ~ 0.3—0.4 $M_{\odot}$, and the stellar XUV flux will be much lower than assumed by the scaling relations used in this paper.